\documentclass{edp-conf}
\usepackage{graphicx}
\begin{document}

\TitreGlobal{SF2A 2003}

\title{A library of galaxy mergers}
\author{Combes, F., Melchior A.-L.}
\address{LERMA, Observatoire de Paris, 61 Av. de l'Observatoire, F-75014, Paris, France}
\runningtitle{GALMER}
\setcounter{page}{237}
\index{Combes, F.}
\index{Melchior, A.-L.}

\maketitle
\begin{abstract} 
We have undertaken a large series of numerical simulations with the goal
to built a library of galaxy mergers (GALMER). Since the aim is to 
have more than a thousand of realisations, each individual run is simplified,
with a small number of particules (24 000), but following
the chemodynamical evolution, with star formation and feedback. We illustrate
in this short report some preliminary results.
\end{abstract}
%

\section{Models and techniques}

 We use a TREE-SPH code to follow the self-gravity
of all components, and the dissipative nature of the gas
(cf Combes \& Melchior 2002).
The 24 000 particles are distributed among stars, gas and
dark matter halo, with varying numbers, according to the
initial morphological types of the galaxies.  
 For this fist series of 128 runs, we have 4 types
of galaxies (i.e. 16 types of initial couples): 
E0, Sa, Sbc and Sd, two masses (giants at 2 10$^{11}$
M$_\odot$ and dwarfs at 5 10$^{10}$ M$_\odot$), two different
vectors for relative initial velocities, and two opposite senses
on these orbits (direct or retrograde). The relative inclination of 
the galaxy planes is fixed to 45$^{\circ}$. The star formation
rate for the various runs is time-averaged in Figure 1.

\begin{figure}[h]
   \centering
\rotatebox{-90}{\includegraphics[width=4cm]{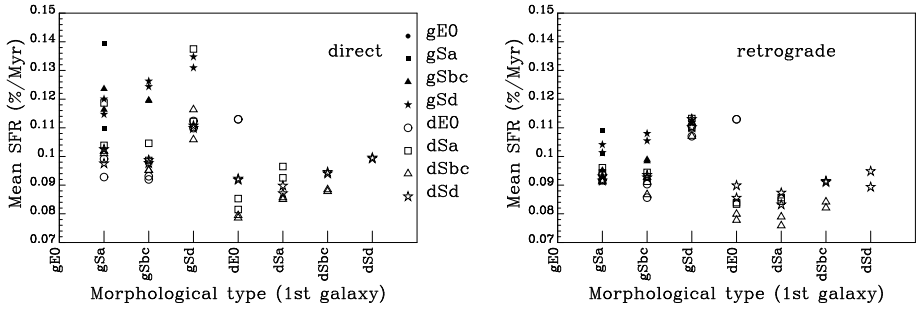}}
      \caption{ Average star formation rate (in units of percentage
of initial gas mass consumed per Myr) over the whole merging simulation
(1.2 Gyr) for various runs as a function of morphological type of the
more massive galaxy of the pair. The type of the second galaxy is indicated
by different symbols. The direct orbits lead in average to
larger SFR.}
       \label{fig1}
   \end{figure}

\section{Star formation recipe}

We have also varied the recipe for the star formation, and tested
the corresponding efficiency. It is basically a ``local'' Schmidt law, where a
fraction of any gas particle transformed to stars
is $\propto \rho^{n-1}$, where $\rho$ is the volumic gas density, 
and $n$ the power of the Schmidt law. An algorithm with hybrid particules is used
(cf Mihos \& Hernquist 1994): when their gas fraction
drops below 5\%, they are then
turned into pure stars, the gas being spread among the neighbours.
The energy of supernovae and stellar winds is partly reinjected in the ISM
under the form of kinetic energy,
through expanding velocity of the surrounding gas.
To simulate the increase of star-formation efficiency in violent
interactions, we also added a term proportional to the local
gas velocity divergence, to a power $\beta$ (this divergence is
counted only when div{\bf v} negative, like in the viscosity term).
 The comparison between the rate of star formation in the merger,
and that in isolated galaxies, depends strongly on the adopted SF rate.
When all gas is consumed before the first pericenter,
the boost due to the dynamical triggering is very limited.
The influence of the div{\bf v} term is less
that the absolute rate of star formation (see figure 2).

\begin{figure}[h]
   \centering
\rotatebox{-90}{\includegraphics[width=5cm]{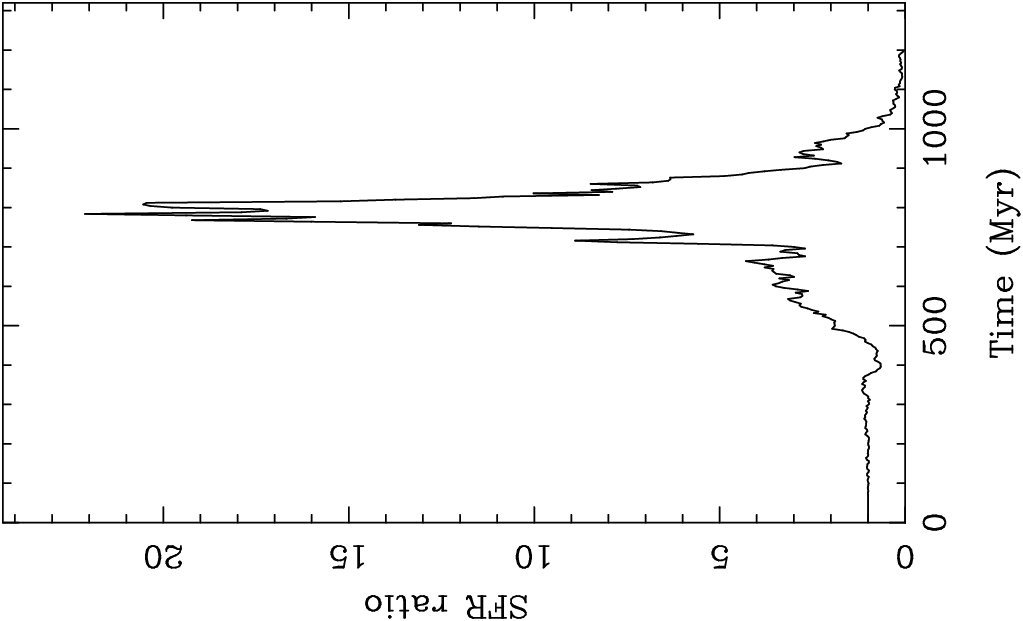}}
\rotatebox{-90}{\includegraphics[width=5cm]{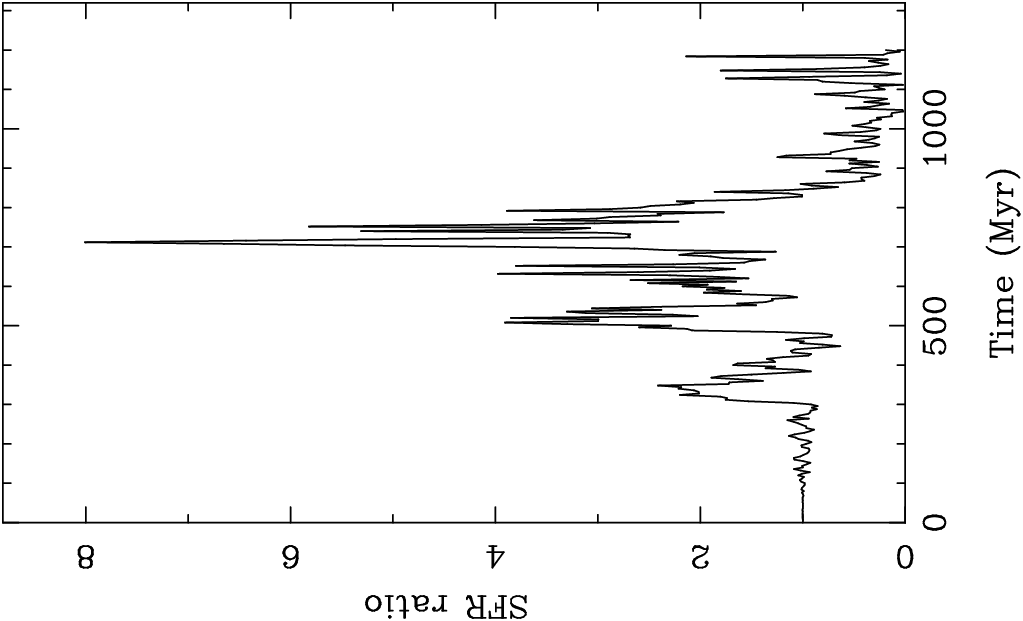}}
      \caption{ Star formation ratio between the merger run and 
the corresponding control run with the two galaxies isolated.
{\bf Left}, the SF recipe is the Schmidt law, with n=1.5;
{\bf Right}, the SF recipe is the same n=1.5 law, but with now a term
proportionnal to the velocity divergence, to simulate the action
of shocks and ISM agitation.}
       \label{fig2}
   \end{figure}


\end{document}